\documentclass[10pt]{article}
\usepackage{amsmath}
\usepackage{amssymb}

\usepackage{graphicx}

\usepackage{cite}

\usepackage{color} 

\usepackage[labelfont=bf,labelsep=period,justification=raggedright]{caption}

\makeatletter
\renewcommand{\@biblabel}[1]{\quad#1.}
\makeatother

\date{\today}

\begin{document}

\begin{flushleft}
{\Large
\textbf{A variational approach to the analysis of non-conservative mechatronic systems}
}


A.~Allison$^{1,\ast}$,
C.~E.~M.~Pearce$^{2}$, 
D.~Abbott$^{1}$
\\
\bf{1} School of Electrical and Electronic Engineering, University of Adelaide, South Australia~5005.\\
\bf{2} School of Mathematical Sciences,                 University of Adelaide, South Australia~5005.\\
$\ast$ E-mail: Corresponding aallison@eleceng.edu.au
\end{flushleft}

\section*{Abstract}
We develop a method for systematically constructing Lagrangian
functions for dissipative mechanical, electrical and, mechatronic
systems. We derive the equations of motion for 
some typical mechatronic systems using
deterministic principles that are strictly variational. We do not
use any {\it ad hoc} features that are added on after the
analysis has been completed, such as the Rayleigh dissipation
function.\\

We generalise the concept of potential, and define
generalised potentials for dissipative lumped system elements. 
Our innovation offers
a unified approach to the analysis of mechatronic systems where there
are energy and power terms in both the mechanical and electrical parts
of the system.  Using our novel technique, we can take advantage of
the analytic approach from mechanics, and we can apply these
powerful analytical
methods to electrical and to mechatronic systems. We can analyse
systems that include non-conservative forces.  
Our methodology is deterministic and does does require any special intuition, and is thus suitable for automation via a computer-based algebra package.\\





\section*{Introduction and motivation}
\label{sec:introduction}

It is a widely believed that the Lagrangian approach
to dynamical systems cannot be applied to dissipative systems that
include non-conservative forces. For example,
Feynman~\cite{feynman_1963} writes that {\it ``The principle of least
  action only works for conservative systems---where all the forces
  can be gotten from a potential function.''}
Lanczos~\cite{lanczos_1949}, writes {\it``Forces of a frictional
  nature, which have no work function, are outside the realm of
  variational principles, while the Newtonian scheme has no difficulty
  in including them. Such forces originate from inter-molecular
  phenomena, which are neglected in the macroscopic description of
  motion. If the macroscopic parameters of a mechanical system are
  completed by the addition of microscopic parameters, forces not
  derivable from a work function would in all probability not
  occur.''}~Lanczos~\cite{lanczos_1949}, 
also writes{ \it ``Frictional forces (viscosity) 
which originate from a transfer of macroscopic into microscopic motions 
demand an increase in the number of degrees of freedom and the application of statistical principles.
They are automatically beyond the macroscopic variational treatment.''}
  It is clear that dissipative forces present a problem to traditional 
  Lagrangian analysis, which means that the Newtonian approach has historically had 
  an advantage, particularly where dissipative forces are significant.\\ 
  
There are a number of formalisms for applying a Newtonian (force-based) approach to mixed mechatronic systems.
The bond-graph approach is based on the systematic use of effort and flow variables.
The work of Karnopp et al.~\cite{karnopp_2000} is important in this regard. We will employ some aspects of Karnopp's work, 
including the homomorphic mappings of variables between different systems. 
There are clear analogies between mechanical and electrical oscillators, and we make use of these.\\

The Newtonian approach has been dominant in practical discipline areas, such as mechanical engineering.
In contrast, the Lagrangian approach, which is very elegant, has
tended to dominate advanced physics texts. For example, the
Hamiltonian approach dominates the subject of quantum mechanics.
Penrose~\cite{penrose_2005}, refers to this paradigm as the {\it ``magical Lagrangian formalism.''} 
He goes on to write that {\it ``The existence of such a mathematically elegant unifying picture appears to be telling us something deep about our physical universe.''}\\

There are a number of more prosaic factors in favour of the Lagrangian approach, which include:
\begin{itemize}

   \item In the Lagrangian formulation, forces of constraint do no work and need not be considered in the analysis. It is often not necessary to calculate internal stresses or forces of reaction.

    \item Post~\cite{post_1962} points out that it is easy to state the underlying physical laws in arbitrary, 
   curvilinear coordinates. It is possible to use generalised coordinates that directly reflect the nature of the physical system.
  
     \item Noether's theorem tells us that, if the Lagrangian function possesses a continuous smooth symmetry, then there will be a conservation law associated with that symmetry~\cite{penrose_2005}. For conservative systems, this leads to the laws of conservation of momentum and conservation of energy. These conservation laws essentially give us one integration of the laws of motion for free. For example we can calculate the final momentum, and the final energy of a system without the need to explicitly integrate the laws of motion.
   
    \item Lagrangian modelling of machines, automatically takes care of energy transfer between different components of a whole system. This prevents incomplete models, which give rise to errors and paradoxes, such as the problem of the Penfield motor~\cite{penfield_1966}. We believe that Lagrangian modelling is a natural choice, where energy is exchanged between different types of storage elements, in such systems as: a moving wire in a magnetic field, the D'Arsonval moving-coil meter, or for mechatronic systems more generally.
   
   \item In the Hamiltonian formulation, only first derivatives are required, not second derivatives.
   
   \item Many quantum systems, such as the hydrogen atom, only have a few degrees of freedom and a 
         complete description of all the microscopic parameters is possible. This means that frictional forces may not even need to be considered.

\end{itemize}

Perhaps the strongest theoretical motivation for the Lagrangian approach is that it explicitly represents the symmetries of the underlying physical laws. 
Melia~\cite{melia_2001} writes: {\it ``As we shall see, the sole motivation for using action principles is to improve our understanding of the underlying physics, with a goal of extracting additional physical laws that might not otherwise be apparent.''}\\

Prior to the work of Riewe~\cite{riewe_1996,riewe_1997}, there was no satisfactory method for completely including non-conservative forces into a variational framework. 
Riewe writes that {\it ``It is a strange paradox that the most advanced methods of classical mechanics deal only with conservative systems, while almost all classical processes observed in the physical world are non-conservative.''} We regard the approach used by Riewe as the most satisfactory method for including non-conservative forces into a variational framework. In this paper we apply his approach, for mechanical systems, to the new areas of electrical and mechatronic systems. We also provide recipes for constructing Lagrangian functions and show, by example, how these techniques can be employed more generally. We believe that the Lagrangian approach naturally models energy exchange within complex machinery, where energy can be stored and transferred between many different forms, including: energy of inertia, elastic energy, frictional loss, energy of the magnetic field, energy of the electric field and resistive loss. Our approach can be used to confer the advantages of the variational method of analysis to a wide range of mechatronic systems, including systems that suffer from dissipative loss.\\  

\subsection*{A short summary of the variational approach}
\label{sec:summary_variational}


We can denote a Lagrangian function for a system as $\mathcal{L}$,
then we can specify the total {\it action} of the system as 
\begin{equation}
\label{eq:action_01}
   \mathcal{I} = \int_{T_1}^{T_2} \mathcal{L} ~d t, 
\end{equation}
where $T_1$ and $T_2$ represent the boundaries of the closed time interval over which we wish to conduct our analysis.
Equation~\ref{eq:action_01} is referred to an {\it action integral.} It is a functional that maps functions, $\mathcal{L}$, onto numbers, $\mathcal{I}$.  
The Euler-Lagrange equation specifies necessary conditions for the first variation of the action integral to vanish, $\delta \left[ \mathcal{I} \right]$ = 0. 
Suppose that the Lagrangian function includes references to a generalised coordinate, $x(t)$, and to its first derivative $\dot{x}$
so $\mathcal{L}= \mathcal{L}\left( x , \dot{x} \right)$, then
the action is extremal when 
we choose $x(t)$ in such a way that the Euler-Lagrange equation is satisfied:
\begin{equation}
\label{eq:euler_lagrange}
   \frac{d}{dt} \left( \frac{\partial \mathcal{L}}{ \partial {\dot{x}} } \right) - \frac{\partial \mathcal{L}}{\partial x}= 0.
\end{equation}
This is the same as saying that all first order variation of the action is zero, $\delta \left[ \mathcal{I} \right]=0$.
The Euler Lagrange equation is an ordinary differential equation that describes the dynamics of the system, in terms of the specified generalised coordinates, such as $x(t)$.

For mechanical systems the Lagrangian is written in terms of energy functions, which are summed together with appropriate sign conventions. They typical symbols are kinetic energy of inertia, $\mathcal{T}\left( {\dot{x}} \right)$, and potential (elastic or gravitational) energy, $\mathcal{V}\left( x \right)$. For these systems the Lagrangian function can be written as: $\mathcal{L}= \mathcal{T} - \mathcal{V}$.  
As we shall see, a classical example is a mass on a spring, where 
$\mathcal{L}= \mathcal{T} - \mathcal{V} = \frac{1}{2} m \dot{x}^2-\frac{1}{2} k x^2$.\\

We will use the notation of Gel'fand~\cite{gelfand_1963}, who denotes a general $k$th order derivative as: $x^{\left( k \right)}$. 
This is more versatile than the more traditional ``dot'' notation, of Newton. 
It is common for Lagrangian analysis to only consider integral derivatives, of low orders, of the generalised coordinates. 
For example, we might consider $x = x^{\left( 0 \right)}$, ${\dot{x}} = x^{\left( 1 \right)}$ and possibly ${\ddot{x}} = x^{\left( 2 \right)}$.
Gel'fand writes the generalised integer-order Euler-Lagrange equation a form that includes higher derivatives, and
is equivalent to:
\begin{equation}
	\label{eq:euler_lagrange_general_integer}
    \sum_{k=0}^{\infty} \left( -1 \right)^{k} \cdot \frac{d^{k}}{ d {t^{k}}} \left(  \frac{\partial \mathcal{L}}{\partial x^{\left( k \right)}}  \right) = 0,
\end{equation}
where where $k \in \mathcal{Z}$, and 
where it is understood that $\frac{\partial \mathcal{L}}{\partial x^{\left( k \right)}} = 0$, for values of $k$ where the Lagrangian has no dependence on  the $k$'th derivative of the coordinate. The proof of Equation~\ref{eq:euler_lagrange_general_integer}, can be obtained by repeatedly integrating by parts, and applying the 
du Bois-Reymond lemma.
Proofs can be found in Gel'fand~\cite{gelfand_1963} and Smith~\cite{smith_1974}.\\


Since the seminal work of Riewe~\cite{riewe_1996,riewe_1997}, a number of other authors have used his approach. These include 
Agrawal~\cite{agrawal_2001}, 
Rabei~\cite{rabei_2004}, 
Frederico~\cite{frederico_2007},
Musielak~\cite{musielak_2007},
Elnabulsi~\cite{elnabulsi_2008} and
Almeida~\cite{almeida_2009,almeida_2010}, but none of these papers have addressed 
the problem of dissipative forces in electrical or mechatronic systems,
which is our main concern here.\\

\subsubsection*{Fractional Calculus}

The indices of differentiation in The Euler Lagrange 
Equation~\ref{eq:euler_lagrange_general_integer} can be fractional, which leads to the formulation:
\begin{equation}
	\label{eq:euler_lagrange_general_fractional}
   \sum_{\forall k} \left( -1 \right)^{k} \cdot \frac{d^{k}}{ d {t^{k}}} \left(  \frac{\partial \mathcal{L}}{\partial x^{\left( k \right)}}  \right) = 0,
\end{equation}
where $k \in \mathcal{Q}$, and
where it is understood that $\frac{\partial \mathcal{L}}{\partial x^{\left( k \right)}} = 0$, for values of $k$ where the Lagrangian has no dependence on  the $k$'th derivative of the coordinate. The proof of this proposition depends on a fractional version of integration by parts, and is found in Riewe~\cite{riewe_1996}.\\

The history of the fractional calculus can be traced back, at least, as far as a paper by Liouville in 1832, where he gives a definition of a fractional derivative:
If a function, $f(t)$ can be written as a sum of exponential functions, or possibly complex-exponential functions, as
\begin{equation}
	\label{eq:exponential_decomposition}
	 f \left( t \right) = \sum_m c_m e^{a_m t},
\end{equation}
for a suitable choice of values, $c_m$ and $a_m$, then we can write the fractional derivative as:
\begin{eqnarray}
\label{eq:liouville_definition}
f^{\left( k \right)}
\frac{d^{k}{f(t)}}{d(t+\infty)^{k}} & = & \frac{d^{k}}{d(t+\infty)^{k}} \sum_m  c_m e^{a_m t} \nonumber \\ 
                                  & = & \sum_m c_m {(a_m)}^k e^{a_m t}.
\end{eqnarray}
Fractional derivatives are not local if $k$ is not an integer, which means that their value depends on a region around the point of evaluation. 
The choice of region is important.
There is also the question as to whether the derivative is causal, or not.
Riemann defined the fractional derivative as:
\begin{equation}
\label{eq:reimann_definition}
f^{\left( -v \right)}
\frac{d^{-v}{f(t)}}{d(t-c)^{-v}} = \frac{1}{\Gamma (v) } \cdot
\int_{T_{1}}^{t} (t-t')^{v-1} f(t') dt'
\end{equation}
provided that $\Re(v)>0$. 
We argue that the derivative is causal if it is a function of the state of the system over the time interval 
that includes all time up to the present moment, $-\infty < t' <t$.
It can be shown that Riemann's definition is consistent with
Liouville's 1832 definition, provided $T_{1} \rightarrow -\infty$, which means that Liouville's definition is causal.
In this paper we use Liouville's definition of fractional derivatives.
This means that we have to express any functions of interest as
sums of exponential functions as indicated in Equation~\ref{eq:exponential_decomposition}. The responses of systems can often be written in this form
using a finite number of terms, so Liouville's definition is very convenient. 
In the more general case, we can use the theory of the Laplace transform to perform the necessary decomposition. In particular, we can write:
\begin{equation}
	\label{eq:laplace_to_derivative}
	\frac{d^{v}{f(t)}}{d t^{v}} =  \textbf{L} ^{-1} \left\{ s^{-v} \textbf{L} \left\{ f \right\}  \left(s \right) \right\}.
\end{equation}
Once again, care has to be taken to integrate the Laplace transform over the correct time interval. This can be achieved by using step-functions as windowing functions, and by making use of shifting theorems. 
The theory of the fractional calculus has been well documented, and summaries can be found in  Oldham~et~al.~\cite{oldham_1974} and briefly in Wiesstein~\cite{weisstein_1999}.\\

\section*{A mechanical harmonic oscillator}
\label{sec:km}

\begin{figure}[!ht]
\begin{center}
\includegraphics[width=0.45\textwidth]{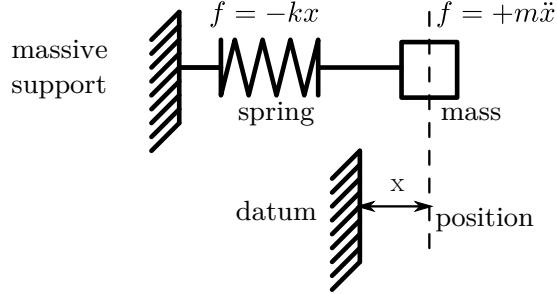}
\end{center}
\caption{{\bf A mass on a spring:~} We consider the simple introductory problem of a mass, $m$, on a spring,
   with Hooke's-law constant of $k$. The kinetic energy stored by the inertia of the mass is denoted by
   $\mathcal{T} = \frac{1}{2} m \dot{x}^2$.
   The elastic potential energy stored in the spring is denoted by
   $\mathcal{V} = \frac{1}{2} k x^2$.
   The independent coordinate is denoted by the position, $x$.
   The Lagrangian function is traditionally written as $\mathcal{L}= \mathcal{T} - \mathcal{V}$, which can be
   written explicitly as $\mathcal{L}=\frac{1}{2} m \dot{x}^2-\frac{1}{2} k x^2$.}
\label{fig:mass_spring_01}
\end{figure}

We consider a common problem from classical mechanics, of a mass on a spring. This problem is widely used to define notation, and can be found in:  
Lamb~ \cite{lamb_1946},
Goldstein~\cite{goldstein_1950},
McCuskey~\cite{mccuskey_1959},
Resnick ~{\&}~Halliday\cite{resnick_1960},
Whylie~\cite{wylie_1960},
Fowles~\cite{fowles_1962},
Feynman~\cite{feynman_1963},
Rabenstein~\cite{rabenstein_1972} and 
Lomen~\cite{lomen_1988},
and many others.

The mechanical harmonic oscillator consists of a mass, spring and massive support (or foundation). 
The complete system is shown in the schematic diagram in Figure~\ref{fig:mass_spring_01}. A mass, $m$, is attached to a spring\footnote{There is some difficulty with the schematic notation for the spring, $k$, since the traditional schematic symbol for a spring resembles the traditional schematic symbol for a resistor. This creates problems if we need to represent both of these different objects in a single drawing. We have followed examples from Giesecke~et al.~\cite{giesecke_1980}. In particular our symbol for the spring has a different aspect ratio to the symbol for the resistor, and the terminations at the ends are different.}, $k$, which is attached to a massive support.
The position of the spring is measured relative to a datum position, which is in a fixed position relative to the massive support. Without any loss of generality we can choose the location of the no-load position of the mass, which gives a simple rule for the stored energy in the spring, $\mathcal{V} = \frac{1}{2} k x^2$.\\ 

The classical problem of a mechanical oscillator is shown in
Figure~\ref{fig:mass_spring_01}. Together the mass and spring form a
mechanical harmonic oscillator. Williams~\cite{williams_1996}
specifies a Lagrangian for this physical system in terms of the single
spatial coordinate, $x$, and writes:
\begin{equation}
\label{eq:mechanical_lagrangian_01}
    \mathcal{L}= \mathcal{T} - \mathcal{V} = \frac{1}{2} m {\dot{x}}^2
                 - \frac{1}{2} k {x}^2,
\end{equation}
where $\mathcal{T}=\frac{1}{2} m \left( \dot{x} \right)^2$, which is
the kinetic energy in the inertia of the mass and
$\mathcal{V}=\frac{1}{2} k \left( x \right)^2$ is the strain energy
stored in the spring.  The Lagrangian function, in
Equation~\ref{eq:mechanical_lagrangian_01} is in a form where we can
directly apply the procedure of Euler and Lagrange Equation to obtain:
\begin{align}
  \label{eq:mechanical_ODE_01}
	m \dot{x} + k x = m x^{(1)} + k x^{(0)} = 0,
\end{align}
 which is the standard Ordinary Differential Equation (ODE) for this system.
Equation~\ref{eq:mechanical_solution_01} can be solved using a number of techniques, 
including the method of the Laplace transform, to obtain:
\begin{equation}
	\label{eq:mechanical_solution_01}
	x(t) = A  \cos \left( \omega t \right) + B \sin \left( \omega t \right),
\end{equation}
where $A$ and $B$ should be chosen in order to satisfy the initial conditions and, $\omega = \sqrt{k/m}$ 
is the un-damped natural angular frequency of oscillation, in radians per second. 
In the completely un-damped case, $\omega$ is also the resonant angular frequency.
We note that the solutions for $x(t)$ can be written as sums of exponential functions, since 
$\cos \left( \theta \right) = \left( e^{+ j \theta}  + e^{- j \theta} \right)/2 $ and 
$\sin \left( \theta \right) = \left( e^{+ j \theta}  - e^{- j \theta} \right)/(2j) $.
This means that Liouville's definition in Equation~\ref{eq:liouville_definition} could be directly applied if we wanted to
determine any fractional derivatives.\\

\section*{A homomorphic mapping}
\label{sec:homomorphic}

The example shown in Figure~\ref{fig:mass_spring_01} is simple and well known, and lies completely within a mechanical problem domain. 
It is not immediately obvious how to extend this type of work to an electrical domain. 
We need a homomorphic mapping of variables that can relate different variables in different physical domains
The mapping needs to relate the names of variables, as well as the set of permissible functions and operators, that work on those variables. 
We use the mapping described in Karnopp et al.~\cite{karnopp_2000}, which is summarised in Table~\ref{tab:homomorphic}.\\

\begin{table*}[!ht]
\caption{
\bf{ Table~~\ref{tab:homomorphic}:}~A homomorphic mapping due to Karnopp et al.}
\begin{tabular}[httb]{|l|l|l|}
\hline \hline     
{\bf Concept}                   & {\bf Mechanical}                               &  {\bf Electrical} \\
\hline \hline     
coordinate                & displacement, $x$                                & charge, $q$ \\
\hline
flow~variable             & velocity, $v = \frac{d x}{d t}$                  & current = $i =  \frac{d q}{d t}$ \\
\hline
energy                    & energy $\mathcal{U}$                             & energy $\mathcal{U}$ \\
\hline
effort~variable           & force, $F=\frac{d \mathcal{U}}{d x}$             & voltage= $v=\frac{d \mathcal{U}}{d q}$ \\
\hline
energy~increment          & $d \mathcal{U} = F \cdot dx$                     & $d \mathcal{U} = v \cdot dq$ \\
\hline
power = $d \mathcal{U} /dt$ & $\mathcal{P}= F \cdot \frac{dx}{dt} = F \cdot v$ & $\mathcal{P}=v \cdot  \frac{dq}{dt} = v \cdot i$\\
\hline
inertial element          & mass, $m$                                        & inductance, $L$ \\
\hline
generalised momentum      & momentum, $p = m \cdot \frac{d x}{d t}$          & magnetic flux, $\Phi = L \cdot \frac{d q}{d t}$\\
\hline
Newton's second law       & $F = m \cdot \frac{d^2 x}{d t^2} = m \cdot a$    &  $v = L \cdot \frac{d^2 q }{d t^2} = L \cdot \frac{d i}{d t}$ \\
\hline
elastic element           & Hooke's law constant, $k$                        &  inverse capacitance, $1/C$ \\
\hline
Hooke's second law        & $F = - k \cdot x$                                &  $v = +\frac{1}{C} \cdot q$ \\
\hline
dissipative element       & damping, $c$                                     &  resistance, $R$ \\
\hline
frictional~force          & $F = - c \cdot \frac{d x}{d t} = -c \cdot v$     & $V = +R \cdot \frac{d q}{d t} = + R \cdot i$ \\
\hline
Joule's law               & $\mathcal{P} = v^2 c$                            & $\mathcal{P} = i^2 R$ \\
\hline
energy of inertia         & $\mathcal{U} = \frac{1}{2} \frac{1}{m} \cdot p^2 =  \frac{1}{2} m \cdot v^2 $ & $\mathcal{U} =  \frac{1}{2} \frac{1}{L} \cdot \Phi^2 =  \frac{1}{2} L \cdot {i}^2 $ \\
\hline
elastic~energy            & $\mathcal{U} = \frac{1}{2} k \cdot x^2 = \frac{1}{2} \frac{1}{k} \cdot F^2 $ & $\mathcal{U} = \frac{1}{2} \cdot \frac{1}{C} \cdot q^2  = \frac{1}{2} C \cdot v^2$ \\
\hline \hline 
\end{tabular}
\begin{flushleft}
{\bf A homomorphic mapping:} The names and purposes of the most important mechatronic dynamical concepts.
\end{flushleft}
\label{tab:homomorphic}
\end{table*}

\section*{An electrical harmonic system}
\label{sec:LC}

\begin{figure}[!ht]
\begin{center}
\includegraphics[width=0.333\textwidth]{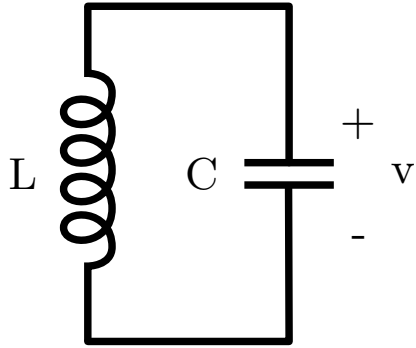}
\end{center}
\caption{{\bf An LC electromagnetic harmonic circuit:~} We consider a capacitor, $C$, in parallel with
   an inductor, $L$. We consider the idealised case where there is no dissipative loss, or resistance $R$.
  The Lagrangian function can be written as $\mathcal{L}= \mathcal{T} - \mathcal{V}$, where
   the magnetic energy stored by the field of the inductor, denoted by
   $\mathcal{T} = L~{\dot{q}}^2 /2$, and
   the electrical potential energy stored in the capacitor is denoted by
   $\mathcal{V} = q^2 / \left( 2 C \right)$, and $q=Cv$ is the coordinate, which we
   interpret as the electrical charge that is transferred through the circuit.}
\label{fig:LC_01}
\end{figure}

If we place a capacitor, $C$ in parallel (and series) with an inductor, $L$, as shown in Figure \ref{fig:LC_01},
then the resulting system will form an electromagnetic harmonic oscillator.\\

If we temporarily ignore the presence of resistance, then we obtain 
the circuit in Figure~\ref{fig:LC_01} is the exact analogue of the
mechanical system in Figure~\ref{fig:mass_spring_01}.  In order to
emphasise the homomorphic mapping between the mechanical and
electrical domains, we map the Lagrangian function in
Equation~\ref{eq:mechanical_lagrangian_01}, using the homomorphic
mapping in Table~\ref{tab:homomorphic}, to obtain\begin{equation}
\label{eq:LC_lagrangian_01}
    \mathcal{L}= \mathcal{T} - \mathcal{V} = \frac{L}{2} {\dot{q}}^2 - \frac{1}{2C}   {q}^2,
\end{equation}
where $L$ is the inductance, $C$ is the capacitance and $q$ is the charge that is transferred through the
circuit.
We note that Equation~\ref{eq:LC_lagrangian_01} is a correct Lagrangian
function for the circuit in Figure~\ref{fig:LC_01}.\\

In Equation~\ref{eq:LC_lagrangian_01}, we use the variable $q$ as a coordinate, in accordance with
homomorphic mapping due to Karnopp. It is the more usual practice in electrical engineering to use the voltage across
a capacitor, $v$, as though it were a generalised coordinate.
Fortunately, it is possible to subject the Lagrangian function in Equation~\ref{eq:LC_lagrangian_01} to a Legendre
transformation, of $q = C v$,
to obtain a new Lagrangian function that is consistent with the previous Lagrangian function (in
terms of energy exchange), but uses the conventional coordinate of $v$ (rather than $q$). 
This new Lagrangian function is self-contained in
the sense that the energy terms for inductor only include references to parameters that pertain to the inductor,
and the energy terms for capacitor only include references to parameters that pertain to the capacitor. There
are no cross-terms. If we impose this last condition then the Legendre transformation is unique and we obtain
a new Lagrangian function:
\begin{equation}
\label{eq:LC_lagrangian_02}
    \mathcal{L}= \mathcal{T} - \mathcal{V} = \frac{1}{2 L} \left( v^{\left( -1 \right)}\right)^2 -
    \frac{C}{2} \left( v^{\left( 0 \right)}
    \right)^2,
\end{equation}
where the independent generalised coordinate is now $v$. The function $v^{\left( 0 \right)}$ is the zeroth derivative of
$v$, which is identical with $v$. We can write $v^{\left( 0 \right)}=v$. The function $v^{\left( -1 \right)}$
denotes the derivative of $v$ to the order of -1, which is equivalent to the integral of $v$. We can write
$v^{\left( -1 \right)} \left( t \right) = \int_{T_1}^{t} v \left( \tau \right) d \tau$.\\

\subsection*{Lagrangian terms for some common lumped mechatronic elements}
\label{sec:lagrangian_list}

We can see from the last example that electrical and mechanical
systems can be mapped onto one another but some care has to be taken
with regard to what we regard as a coordinate. The canonical choice
for a massive particle is to regard the spatial position as the coordinate
and to regard the generalised momentum as the other variable of
interest. These choices are not arbitrary. The coordinate, ${\bf{x}}$, must be an
exact differential, for example: $\oint \bf{x} d{\bf{x}} = 0$ for all
possible closed paths. In quantum mechanics, position, ${\bf{x}}$, and
momentum, ${\bf{p}}$, are conjugate variables.  The relationship between
${\bf{x}}$ and ${\bf{p}}$ is a physical phenomenon, not just an arbitrary
choice.  Finally we know from classical mechanics that ${\bf{x}}$
and ${\bf{p}}$ play a role in Liouville's theorem. See Reif~\cite{reif_1965}
and Wannier~\cite{wannier_1966}, for example. Liouville's theorem would not
apply in the same way if we were to describe particle motion in terms of
force and velocity, rather than position and momentum. 
If we want to
apply Liouville's theorem to a complicated electrical system with many
degrees of freedom then we really need to use magnetic flux,
${\bf{\Phi}}$, and electric flux, ${\bf{q}}$, to describe each element of
the system. This has been carried out in some specialist areas, such as
Allison~\cite{allison_2009}, but it is not common and is not likely to become 
universal in the electrical engineering literature in the foreseeable future.\\

The conventional choice of electrical variables, voltage, $v$, and
current, $i$ are of a fundamentally different type to the conventional
choice of mechanical variables, ${\bf{x}}$ and ${\bf{p}}$.  These
incompatible conventions are not likely to change.  The best solution
seems to be that we should re-write the Lagrangian terms, using the
conventional electrical variables, but to do so in such a way that the
energy values are preserved, and all sign and phase relationships are
consistent. These conversions are not physical laws.  They have the
same status as the conversion from degrees Fahrenheit to Kelvin, for example. 
 The Lagrangian
terms, using common electrical engineering variables, are shown in
Figures~\ref{tab:lagrangian_term_current}~and~\ref{tab:lagrangian_term_voltage}.
If we wish to include one of these devices in a system, and we wish to
carry out Lagrangian analysis then it is only necessary to look up the
term in the table and to include it in the appropriate way, in the
Lagrangian function.  No other special modelling or cognition is
required.  The process is direct enough to be able to be carried out
by computer.\\

\begin{table*}[!ht]
   \caption{
   \bf{Table of Lagrangian terms, in terms of current}}
      \begin{tabular}[httb]{|l|l|l|l|l|}
         \hline \hline     
            {\bf parameter} & {\bf phasor} & {\bf Lagrangian} $\mathcal{L}$ &  {\bf order}, $k$ & {\bf Euler-Lagrange}\\
            \hline \hline     
            $L$ & $I = \left( -j / \left( \omega L \right) \right) V$ & $\frac{1}{2} L \left( i^{\left( 0 \right)} \right)^2 $ & $0$ & $+L i^{\left( 0 \right)}$ \\ 
            \hline
            $R$ & $I = \left( 1 / R \right) V$  & $\frac{+j}{2} R \left( i^{\left( -1/2 \right)}\right)^2$ & $-1/2$ &  $+R i^{\left( -1 \right)}$ \\
            \hline
            $C$ & $I = \left( +j \omega C \right) V$ & $\frac{-1}{2 C} \left( i^{\left( -1 \right)}\right)^2$ & $-1$ & $+\frac{1}{C} i^{\left( -2 \right)}$ \\ 
            \hline \hline 
   \end{tabular}
   \begin{flushleft}
      {{\bf Lagrangian terms, with current:}
      We list the common electrical lumped parameters,
      and compare the admittance with the corresponding term from the Lagrangian function.  
      We also list the order of differentiation, $k$, and the corresponding term from the Euler-Lagrange equation.
      The phase direction of the Lagrangian term leads the phase direction of the admittance by $90^{\circ}$. 
      This is equivalent to multiplying the Lagrangian term by $+j$. We can multiply the Lagrangian term by any constant that we like,
      as long as we do this consistently.  If we were to remove the factor of $+j$ then the Lagrangian terms and the admittances 
      would have consistent phases, but all the Lagrangians would have new phases and some of these would not be consistent with existing practice in 
      mechanics. In this paper, we rigorously adopt the convention that is used in mechanics, which means that we do not use the sign convention 
      that is common in electrical engineering.}
   \end{flushleft}
   \label{tab:lagrangian_term_current}
\end{table*}
\begin{table*}[!ht]
   \caption{{\bf Table of Lagrangian terms, in terms of voltage}, $v$;~Table~\ref{tab:lagrangian_term_voltage}}
   \begin{tabular}[httb]{|l|l|l|l|l|}
      \hline \hline     
      {\bf parameter} & {\bf Phasor} & {\bf Lagrangian} $\mathcal{L}$ &  {\bf order}, $k$ & {\bf Euler-Lagrange}\\
      \hline \hline     
      $L$ & $V = \left( j \omega L \right) I$ & $\frac{1}{2 L}  \left( v^{\left( -1 \right)}\right)^2$   & $-1$ & $-\frac{1}{L} v^{\left( -2 \right)}$ \\ 
      \hline
      $R$ & $V = R  I$  & $\frac{-j}{2R} \left( v^{\left( -1/2 \right)}\right)^2$ & $-1/2$ &  $-\frac{1}{R} v^{\left( -1 \right)}$ \\
      \hline
      $C$ & $V = {-j}{\omega C}  I$ & $\frac{-C}{2} \left( v^{\left( 0 \right)}\right)^2$   & $0$ & $-C v^{\left( 0 \right)}$ \\ 
      \hline \hline 
   \end{tabular}
   \begin{flushleft}
     {{\bf Lagrangian terms, with voltage:}
     We list common lumped electrical parameters,
     and compare the impedance with the corresponding term from the Lagrangian function.  
     We also list the order of differentiation, $k$, and the corresponding term from the Euler-Lagrange equation.
     The phase direction of the Lagrangian term lags the phase direction of the impedance by $90^{\circ}$. 
     This is equivalent to multiplying the Lagrangian term by $-j$. We can multiply the Lagrangian term by any constant that we like,
     as long as we do this consistently.  If we were to remove the factor of $-j$ then the Lagrangian terms and the admittances 
     would have consistent phases, but all the Lagrangians would have new phases and some of these would not be consistent with existing practice in 
     mechanics. In this paper, we rigorously adopt the convention that is used in mechanics, which means that we do not use the sign convention 
     that is common in electrical engineering.}
   \end{flushleft}
   \label{tab:lagrangian_term_voltage}
\end{table*}

In the interests of consistency, we include a table for the more familiar Lagrangian terms, in Table~\ref{tab:lagrangian_term_mechanical}.
\begin{table*}[!ht]
   \caption{{\bf Table of mechanical Lagrangian terms,}~Table~\ref{tab:lagrangian_term_mechanical}:}
   \begin{tabular}[httb]{|l|l|l|l|l|}
      \hline \hline     
      {\bf parameter} & {\bf Lagrangian~3D} & {\bf Lagrangian~1D} &  {\bf order}, $k$ & {\bf Euler-Lagrange}\\
      \hline \hline     
      $m$ & $\frac{1}{2m} \left| {\bf p}\right|^2 $ & $\frac{1}{2} m \left( x^{\left( 1 \right)} \right)^2$ & $1$ & $-m x^{\left( 2 \right)}$ \\
      \hline 
      $c$ & \texttt{not applicable} & $+\frac{j}{2} c \left( x^{\left( 1/2 \right)} \right)^2 $ & $1/2$ & $-c x^{\left( 1 \right)}$ \\
      \hline 
      $k$ & $-\frac{1}{2} k \left| {\bf x}\right|^2 $ & $-\frac{1}{2} k \left( x^{\left( 0 \right)} \right)^2$ & $0$ & $-k x^{\left( 0 \right)}$ \\  
      \hline \hline 
   \end{tabular}
   \begin{flushleft}
      {{\bf Lagrangian terms for mechanical parameters:}
      We list the common lumped mechanical parameters. From left to right, we list the common symbol for the parameter,
      the Lagrangian for the 3D vector case (in terms of position or momentum), the Lagrangian for the 1D case (in terms of the position only),
      the order of differentiation employed and the resulting them in the Euler-Lagrange Equation. 
      We use the sign convention of {Gel{'}fand}~and~Fomin~\cite{gelfand_1963}
      for the terms of the Euler-Lagrange Equation. Some authors introduce an additional minus sign, in order to make all of these terms positive.}
   \end{flushleft}
   \label{tab:lagrangian_term_mechanical}
\end{table*}

Finally, we note that some of the Lagrangian terms are imaginary and
that the resulting Lagrangian functions will, in general, be complex.
We consider arithmetic operations to operate within the field of
complex numbers, $\mathcal{C}$.  The traditional case, of real
Lagrangian functions, is a special case of our more general
formulation.  Our formulation is consistent with the earlier work of
Illert~\cite{illert_1989}, who applied the concept of complex
Lagrangian functions to the classical seashell problem.\\

\section*{A damped mechanical harmonic system}
\label{sec:kcm}
\begin{figure}[!ht]
\begin{center}
\includegraphics[width=0.45\textwidth]{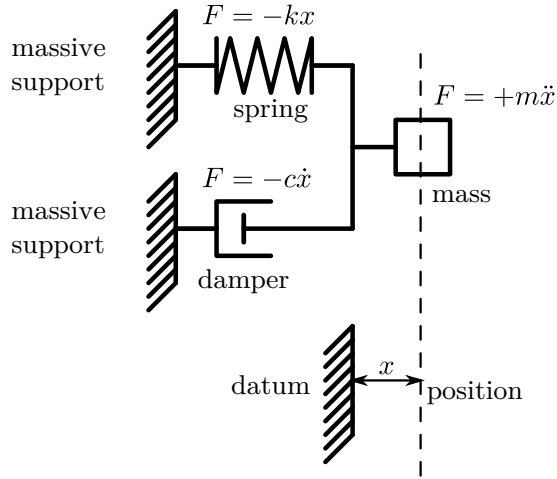}
\end{center}
\caption{{\bf A damped mechanical harmonic oscillator:~} By introducing non-conservative, or dissipative,
elements into a system we need to generalise our concept of potential. The Lagrangian for this system can be
written as $\mathcal{L} = \frac{1}{2} m \left( x^{\left( 1 \right)} \right)^2 - \frac{1}{2} k \left( x^{\left( 0
\right)} \right)^2 + \frac{j}{2} c \left( x^{\left( 1/2 \right)} \right)^2$. The additional term differs from
the terms for the un-damped system in two key ways: the term is complex has an imaginary phase, of $+j$, and
there is a fractional derivative of the coordinate, $x^{\left( 1/2 \right)}$.}
\label{fig:kcm}
\end{figure}

We consider the damped mechanical oscillator, as shown in Figure~\ref{fig:kcm}. 
This problem is solved by Riewe~\cite{riewe_1996}. In our case, we only need to take the 
Lagrangian terms from Table~\ref{tab:lagrangian_term_mechanical} and add them together 
to form the Lagrangian function for the system. This is shown as follows,
\begin{eqnarray}
\label{eq:kcm_lagrangian} \mathcal{L} 
& = &   \frac{1}{2} m \left( \dot{x} \right)^2 
      + \frac{j}{2} c \left( x^{\left( 1/2 \right)} \right)^2
      -  \frac{1}{2} k \left( x \right)^2 \nonumber \\
& =  &  \frac{1}{2} m \left( x^{\left( 1 \right)} \right)^2 
      + \frac{j}{2} c \left( x^{\left( 1/2 \right)} \right)^2
      - \frac{1}{2} k \left( x^{\left( 0 \right)} \right)^2.
\end{eqnarray}
The resulting Euler-Lagrange equation can be assembled from the Euler-Lagrange terms in 
Table~\ref{tab:lagrangian_term_mechanical}, or calculated directly, using
Equation~\ref{eq:euler_lagrange_general_fractional}. The result is given by
\begin{equation}
  \label{eq:damped_mechanical_harmonic_euler_lagrange}
  - \left(  m x^{\left( 2 \right)} +c x^{\left( 1 \right)} +k x^{\left( 0 \right)}  \right) = 0, 
\end{equation}
which is the same result that we would obtain if we used a free body diagram and Newton's laws of motion.\\
\section*{The use of constraints}
\label{sec:constraints}
The use of the calculus of variations to evaluate extremal functions,
subject to constraints is described in a number of references,
including Lanczos~\cite{lanczos_1949}. It is possible to regard
perfect sources, of voltage or charge (or force or velocity), as
constraints. This can simplify the working of some problems.  We
include an example here.\\ 

We consider the case of a purely 
resistive system.
Jaynes~\cite{jaynes_1980} traces this problem
back as far as Kirchhoff~\cite{kirchhoff_1848}, and points out that the
condition that no electric charge should accumulate at any point in a
resistive material requires that $\nabla \cdot \left( \sigma \nabla
v(x) \right) =0$, which is just the Euler-Lagrange equation stating
that the production of Joule heat in a domain
 $\mathcal{D}$, 
$\int_{\mathcal{D}} \sigma \left( \nabla v \right)^2 d \mathcal{D}$ 
is stationary with respect to
variations $\delta {v}\left( x\right)$ that vanish at the boundary
of $\mathcal{D}$.\\

We also note that Kirchhoff's voltage law is a tautology based on the
definitions of energy and voltage.  In a quasi-static situation where
radiation is not significant voltage is just energy per unit charge,
$v = \Delta U / \delta q$.  It is tempting to regard Kirchhoff's
current law as a statement of the conservation of charge, but this is
misleading.  Even if we grant the continuity of charge, $\nabla J
+ \partial \rho / \partial t = 0$ then it would still be possible to
have accumulation of charge.  The equivalent principles of ``no
accumulation of charge'' and ``minimum production of Joule heat in a
domain'' are ultimately statistical in nature and are related to the
second law of thermodynamics. This is discussed in
Allison~\cite{allison_2009}.\\

In this paper, we consider the special case where the parameters are lumped into 
two resistors in series, as shown in
Figure~\ref{fig:2R}. 
To simplify the notation, we define a gradient operator, $\nabla$, over variations with respect to the variational operators,
$\delta_{v_1}$ and $\delta_{v_2}$, rather than partial derivatives, 
$\partial / \partial v_1 $ and $\partial / \partial v_2 $. We can define
\begin{equation}
  \label{eq:del_definition}
  \nabla \mathcal{L} = \left[
    \begin{array}{c}
      \delta_{v_1} \mathcal{L} \\
      \delta_{v_2} \mathcal{L}
    \end{array}
\right].
\end{equation}
This allows us to use the notation of the Lagrange multiplier to write down a necessary condition for a constrained optimum.
If the independent variables, $v_1$ and $v_2$ are constrained by a function of constraint 
\begin{equation}
  \label{eq:constraint_function}
   \Psi \left( \left[
    \begin{array}{c}
      {v_1} \\
      {v_2}
    \end{array}
\right] \right) = {\bf 0},
\end{equation}
then we can only obtain constrained stationary values of $\mathcal{L}$ when
\begin{equation}
  \label{eq:lagrange_multiplier}
  \nabla \mathcal{L} - \lambda \nabla \Psi = 0.
\end{equation}
We apply this principle to the problem in the next section.\\

\section*{A problem with two resistors}
\label{sec:2R}

\begin{figure}[!ht]
\begin{center}
\includegraphics[width=0.45\textwidth]{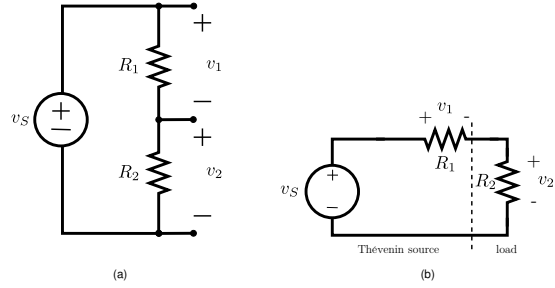}
\end{center}
\caption{{\bf Two resistors in series:~} The voltage source, $v_S$ places a
   constraint on the voltages across the two resistors, $v_1$ and $v_2$. Kirchhoff's voltage law
   implies that $\Psi = v_1 + v_2 - v_S = 0$. We can regard the function $\Psi$ as a function of constraint.
   Kirchhoff used a function that is equivalent to the dissipated power as a Lagrangian function.
   In modern notation we can write
   $\mathcal{L}_K = P = \left( v_1 \right)^2/R_1 + \left( v_2 \right)^2/R_2$. In order to ensure compatibility
   with other existing Lagrangian functions, we need to apply a transformation to obtain a new Lagrangian function of
   $\mathcal{L} = \frac{-j}{2 R_1} \left( {v_1}^{\left(-1/2\right)} \right)^2 +
   \frac{-j}{2 R_2} \left( {v_2}^{\left(-1/2 \right)} \right)^2$.
   This Lagrangian function has been multiplied by a scalar of $-j/2$ and  the order of differentiation has been reduced
   from $v^{\left( 0 \right)}$ to $v^{\left( -1/2 \right)}$, which is equivalent to a half-integration of the
   Lagrangian function, or a full integration of the resulting Euler Lagrange equation.}
\label{fig:2R}
\end{figure}

In Figure~\ref{fig:2R}(a), the voltage source, $v_S$ places a constraint on the voltages across the two resistors,
$v_1$ and $v_2$. We can apply Kirchhoff's voltage law to the single mesh in this circuit to obtain $\Psi = v_1 +
v_2 - v_S = 0$, where $\Psi$ is a function of constraint. Gradients of this constraint function are needed in
order to determine the constrained stationary functions of the system. Kirchhoff noted that the voltages in a
resistive circuit, such as $v_1$ and $v_2$, would arrange themselves in such a way as to minimise the dissipated
heat energy, given by Joule's law. Jaynes points out that this is equivalent to defining a ``Kirchhoff''
Lagrangian function. For us, this takes the form $\mathcal{L}_K =$ power $= {v_1}^2/R_1 + {v_2}^2/R_2$. 
In order to ensure compatibility with other existing Lagrangian functions, we apply a Legendre
transformation to obtain a new Lagrangian function
\begin{equation}\label{eq:2R}
\mathcal{L} = \frac{-j}{2 R_1} \left( {v_1}^{\left(-1/2\right)} \right)^2 +
   \frac{-j}{2 R_2} \left( {v_2}^{\left(-1/2\right)} \right)^2.
\end{equation}
The resulting Lagrangian function, in Equation~\ref{eq:2R}, can also
be assembled from the Euler-Lagrange terms in
Table~\ref{tab:lagrangian_term_voltage}. It is consistent.
In Figure~\ref{fig:2R}(b) we indicate that this situation is very common since it occurs whenever a linear source is
connected to a resistive load.\\

We can use the principle of the Lagrange multiplier to obtain equations for the
stationary functions, subject to constraints.
We begin by calculating the individual variations:
$\delta_{v_1} \mathcal{L} = - \left( 1 / R_1 \right) v_1^{\left( -1 \right)}$, and
$\delta_{v_2} \mathcal{L} = - \left( 1 / R_2 \right) v_2^{\left( -1 \right)}$, which leads to the following 
form for the gradient, $\nabla \mathcal{L}$, as
\begin{equation}
  \label{eq:del_L_R2}
  \nabla \mathcal{L} = \left[
    \begin{array}{c}
      \delta_{v_1} \mathcal{L} \\
      \delta_{v_2} \mathcal{L} \\
    \end{array}
     \right]
  = \left[
    \begin{array}{c}
      -\frac{1}{R_1} {v_1}^{\left( -1 \right)} \\
      -\frac{1}{R_2} {v_2}^{\left( -1 \right)} \\
    \end{array}
\right].
\end{equation}
The function of constraint is $\Psi = v_1 + v_2 - v_S = 0$, and we obtain the gradient of this as
\begin{equation}
  \label{eq:del_L_constraint_R2}
  \nabla \Psi = \left[
    \begin{array}{c}
      \delta_{v_1} {\Psi} \\
      \delta_{v_2} {\Psi} \\
    \end{array}
    \right]
  = \left[
    \begin{array}{c}
      1 \\
      1 \\
    \end{array}
\right].
\end{equation}
We can apply the principle of the Lagrange-multiplier to obtain
\begin{equation}
  \label{eq:lagrange_multiplier_R2}
   \left[
      \begin{array}{c}
         -\frac{1}{R_1} {v_1}^{\left( -1 \right)} \\
         -\frac{1}{R_2} {v_2}^{\left( -1 \right)} \\
      \end{array}
   \right]
   - \lambda
   \left[
      \begin{array}{c}
         1 \\
         1 \\
      \end{array}
   \right] =0,  
\end{equation}
for some constant complex number, $\lambda \in \mathcal{C}$.
If we differentiate once, with respect to time, and multiply by $-1$, we obtain:
\begin{equation}
  \label{eq:solution_R2}
  i = \frac{v_1}{R_1} = \frac{v_2}{R_2}, 
\end{equation}
where $i \in \mathcal{C}$ is a complex number. Since $v_1$, $R_1$, $v_2$ and $R_2$ are all real
we an infer that $i$ is real. If we consider Ohm's law then $i$ is a common current that is shared by both 
resistors. Equation~\ref{eq:solution_R2} can also be obtained by using Kirchhoff's current law and by applying 
Ohm's law twice. Our main aims in presenting this last example are:
\begin{itemize}
   \item to illustrate the use of constraints, with possible time and rate dependence,
   \item to demonstrate the utility of our extended Lagrange-multiplier notation,
   \item to provide a historical reference to the important work of Kirchhoff and Jaynes, and
   \item to resolve an apparent contradiction between the Lagrangian analysis of purely reactive systems
         which only have energy storage, and the Lagrangian analysis of purely resistive systems,
         which only have power dissipation. The reactive systems have Lagrangian terms that only depend on energy terms.
         The resistive systems have Lagrangian terms that only depend on power terms. Superficially, this appears to be a contradiction.
\end{itemize}

In order to examine the proper relationship between ``energy'' Lagrangian terms and ``power'' Lagrangian terms further, we 
next consider a mixed example, where the system contains both resistive and reactive elements, tied together with a constraint.\\

\section*{A damped electrical harmonic system}
\label{sec:LRC}

\begin{figure}[!ht]
\begin{center}
\includegraphics[width=0.45\textwidth]{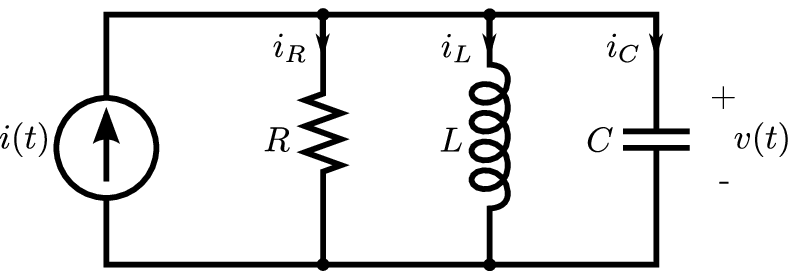}
\end{center}
\caption{{\bf A parallel RLC circuit, with source:~}
We can use the established rules to write the Lagrangian function for the circuit as:
$\mathcal{L} = +\left({j R}/{2} \right)  \left( {i_R}^{\left(-1/2\right)} \right)^2
               +\left({L}/{2}\right)  \left( {i_L}^{\left( 0 \right)} \right)^2
              -\left({1}/\left({2C}\right)\right) \left( {i_C}^{\left( -1 \right)} \right)^2.$
We also use Kirchhoff's current law to impose a
constraint function of $\Psi = i_R + i_L + i_C - i_S = 0$, and we use the principle of the Lagrange multiplier to 
obtain an ordinary differential equation to describe the dynamical behaviour of $v(t)$.}
\label{fig:LRC}
\end{figure}

We can use the terms in Table~\ref{tab:lagrangian_term_current}
to write the Lagrangian function for the circuit 
in Figure~\ref{fig:LRC}
as:
\begin{equation}\label{eq:LRC_lagrangian}
\mathcal{L} = +\frac{j}{2} R \left( {i_R}^{\left(-1/2\right)} \right)^2
              +\frac{1}{2} L \left( {i_L}^{\left(0 \right)} \right)^2
              -\frac{1}{2C} \left( {i_C}^{\left( -1 \right)} \right)^2.
\end{equation}
We can use Kirchhoff's current law to impose a
constraint function of $\Psi = i_R + i_L + i_C - i_S = 0$. 
We can then use the techniques from the last section to
obtain the solution to the problem of the dynamics
of this circuit. 
We then find a functions, $v(t)$, 
that give stationary values for the of the action, $\mathcal{I}$, subject to the constraint,
$\Psi$. The principle of the Lagrange multiplier allows us to replace the optimising principle with a new
constraint, 
\begin{equation}
\label{eq:Lagrange_multiplier_LRC}
 {\nabla_{v}}\left[ \mathcal{J} \right] - \lambda \cdot {\nabla_{v}}\left[ \Psi \right] = 0. 
\end{equation}
We can
combine this new constraint with the original constraint, $\Psi=0$, and use algebraic techniques to obtain an
ordinary differential equation for the dynamics of the circuit.\\

We can use the Euler-Lagrange terms in Table~\ref{tab:lagrangian_term_current} to write:
\begin{equation}
  \label{eq:del_L_LRC}
  \nabla \mathcal{L} = \left[
    \begin{array}{c}
      \delta_{i_R} \mathcal{L} \\
      \delta_{i_L} \mathcal{L} \\
      \delta_{i_C} \mathcal{L} \\
    \end{array}
     \right]
  = \left[
    \begin{array}{c}
      +R           {i_R}^{\left( -1 \right)} \\
      +L           {i_L}^{\left(  0 \right)} \\
      +\frac{1}{C} {i_C}^{\left( -2 \right)} \\
    \end{array}
\right].
\end{equation}
The function of constraint is derived from Kirchhoff's current law
and can be written as
$\Psi = i_R + i_L + i_C - i_S$, and we obtain the gradient of this as
\begin{equation}
  \label{eq:del_L_constraint_LRC}
  \nabla \Psi = \left[
    \begin{array}{c}
      \delta_{i_R} {\Psi} \\
      \delta_{i_L} {\Psi} \\
      \delta_{i_C} {\Psi} \\
    \end{array}
    \right]
  = \left[
    \begin{array}{c}
      1 \\
      1 \\
      1 \\
    \end{array}
\right].
\end{equation}
We can apply the principle of the Lagrange-multiplier to obtain
\begin{equation}
  \label{eq:lagrange_multiplier_LRC}
\left[
    \begin{array}{c}
      +R           \cdot {i_R}^{\left( -1 \right)} \\
      +L           \cdot {i_L}^{\left(  0 \right)} \\
      +\frac{1}{C} \cdot {i_C}^{\left( -2 \right)} \\
    \end{array}
\right]
- \lambda \cdot 
\left[
    \begin{array}{c}
      1 \\
      1 \\
      1 \\
    \end{array}
\right]
=0, 
\end{equation}
for some constant complex number, $\lambda \in \mathcal{C}$,
which leads to the result
\begin{equation}
  \label{eq:LRC_result}
  +R \cdot {i_R}^{\left( -1 \right)} = +L \cdot {i_L}^{\left(  0 \right)} = +\frac{1}{C} \cdot {i_C}^{\left( -2 \right)} = \lambda.
\end{equation}
If we apply the constitutive laws for the three devices, and
Kirchhoff's voltage law, three times, then we realise that we can
interpret $\lambda$ as the common voltage across all three components,
$\lambda = v$.  Of course, we could have obtained this result using
more conventional circuit theory but the point here is that we have
arrived at differential equations for the system 
in Figure~\ref{fig:LRC}
using purely
variational techniques and we have been able to model a complete electrical
system that includes a dissipative element, $R$.\\

\section*{A ladder filter}
\label{sec:RCRCR}

\begin{figure}[!ht]
\begin{center}
\includegraphics[width=0.45\textwidth]{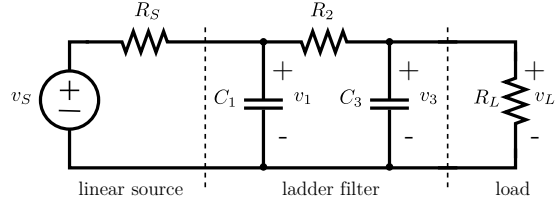}
\end{center}
\caption{{\bf An electrical ladder-filter circuit:~} In this circuit, we make  a careful choice of
   generalised coordinates, which allows us to avoid the explicit use of functions of constraint.
   We can use the established rules to write down the Lagrangian function for the circuit as shown in
   Equation~\ref{eq:RCRCR}.
   We apply the rules for the Euler-Lagrange equation to this Lagrangian function to obtain a pair
   of ordinary differential equations that describe the dynamics of this circuit.}
\label{fig:RCRCR}
\end{figure}

The use of constraints can be a powerful technique, but it does add
some extra complication to the analysis.  It is often possible to make
a careful choice of generalised coordinates, and avoid the need for
constraints.  We demonstrate this concept by analysing the ladder
circuit in Figure~\ref{fig:RCRCR}.\\

If circuits have obvious regularity or symmetry, like the ladder
circuit, then it is often possible to choose the coordinates in such a
way that constraints are automatically obeyed. For the circuit in
Figure~\ref{fig:RCRCR}, we can choose the state-variables as $v_1$ and
$v_3$. The conventional expression for the stored energy in the system
can be written entirely in terms of these state variables as
$\mathcal{U} = C_1 \left(v_1\right)^2 / 2 + C_2 \left(v_2\right)^2 /
2$.  It is also possible to express all the voltages across the
resistors in terms of $v_1$, $v_2$ and $v_s$, using Kirchhoff's
voltage law\footnote{ Kirchhoff's Voltage Law (KVL) does impose
constraints on the system.  We implicitly use KVL, and apply constraints, 
in the definition of the Lagrangian function, $\mathcal{L}$.
This means that we do not need to explicitly use a Lagrange multiplier technique.}.
Our decision to use the state-variables as the
independent coordinates of the system means that the Lagrangian
function in Equation~\ref{eq:RCRCR} takes a simple form and can be
written down almost as quickly as the circuit can be drawn. Our choice
also ensures that the final Euler-Lagrange equations are closely
related to the state variable model, which could be obtained by using
conventional circuit analysis. It is also clear, from the symmetry of
ladder circuits, that we could extend this Lagrangian technique to
ladders of arbitrary length and composition, as long as they were
composed of components from
Table~\ref{tab:lagrangian_term_voltage}.
The Lagrangian function can be written directly as:
\begin{eqnarray}\label{eq:RCRCR}
\mathcal{L} & = & - \frac{j}{2 R_S} \left( \left( v_1 - v_S \right)^{\left( -1/2 \right)} \right)^2 \nonumber \\ 
            &   & - \frac{1}{2} C_1 \left( {v_1}^{\left( 0 \right)} \right)^2 \nonumber \\ 
            &   & - \frac{j}{2 R_2} \left( \left( v_3 - v_1 \right)^{\left( -1/2 \right)} \right)^2 \nonumber \\
            &   & - \frac{1}{2} C_3 \left( {v_3}^{\left( 0 \right)} \right)^2 \nonumber \\
            &   & - \frac{j}{2 R_L} \left( {v_3}^{\left( -1/2 \right)} \right)^2.
\end{eqnarray}
This Lagrangian function is completely composed of terms that can be
found in Table~\ref{tab:lagrangian_term_voltage}.
We can use the established rules to calculate the variations of the Lagrangian:
\begin{eqnarray}
  \label{eq:variation_01_RCRCR}
  \delta_{v_1} \mathcal{L} & = & -C_1 v_1^{\left( 0 \right)} \nonumber \\
                          &   & -\frac{1}{R_S} \left( v_1^{\left( -1 \right)}  - v_S^{\left( -1 \right)} \right) \nonumber \\
                          &   & -\frac{1}{R_2} \left( v_1^{\left( -1 \right)}  - v_3^{\left( -1 \right)} \right)=0 
\end{eqnarray}
and
\begin{eqnarray}
  \label{eq:variation_02_RCRCR}
  \delta_{v_3} \mathcal{L} & = & -C_3 v_3^{\left( 0 \right)}  \nonumber \\
                          &   & -\frac{1}{R_2} \left( v_3^{\left( -1 \right)}  - v_1^{\left( -1 \right)} \right) \nonumber \\
                          &   & -\frac{1}{R_L} \left( v_3^{\left( -1 \right)} \right)=0.
\end{eqnarray}
We can differentiate these variations, once with respect to time and rearrange
the equations into the usual form of a state-variable model of the form 
\begin{equation}
  \label{eq:variation_matrix_RCRCR_00}
  {\dot{\mathbf{V}}} = \mathbf{A} \mathbf{V} + \mathbf{B},
\end{equation}
where the time-rate of change is
\begin{equation}
  \label{eq:variation_matrix_RCRCR_01}
 {\dot{\mathbf{V}}} = \frac{d}{dt}
\left[
    \begin{array}{c}
       v_1 \\
       v_3 \\
    \end{array}
\right], \nonumber
\end{equation}
and the transition-matrix is
\begin{equation}
  \label{eq:variation_matrix_RCRCR_02}
\mathbf{A} = \left[
    \begin{array}{cc}
      -\left( \frac{1}{R_s} + \frac{1}{R_2} \right) \frac{1}{C_1} & \frac{+1}{R_2 C_1} \\
       \frac{+1}{R_2 C_3}                                         & -\left( \frac{1}{R_2} + \frac{1}{R_L} \right)\cdot \frac{1}{C_3} \\
    \end{array}
\right], \nonumber
\end{equation}
and the state-vector is
\begin{equation}
\label{eq:variation_matrix_RCRCR_03}
\mathbf{V} = \left[
    \begin{array}{c}
       v_1 \\
       v_3 \\
    \end{array}
\right],  \nonumber
\end{equation}
and the source-vector is
\begin{equation}
\label{eq:variation_matrix_RCRCR_04}
\mathbf{B} = \left[
    \begin{array}{c}
       \frac{1}{R_s C_1} \\
       0         \\
    \end{array}
\right]  v_S.  \nonumber
\end{equation}

There is great advantage in noting the symmetries of any circuit that
is being analysed, and choosing the generalised coordinates in a
consistent manner.  For example, we can exploit the symmetry of
ladder circuits to extend our Lagrangian technique to ladder circuits
of arbitrary length.\\

\section*{A mechatronic problem, the D'Arsonval galvanometer }
\label{sec:darsonval}

One of the great advantages of the Lagrangian approach is that it can be easily used to model 
devices that transduce energy between different forms. For example, an electric motor (or generator) transduces 
energy between electrical energy (in electric and magnetic forms) to, and from, mechanical energy
(in kinetic and elastic forms).

\begin{figure}[!ht]
\begin{center}
\includegraphics[width=0.45\textwidth]{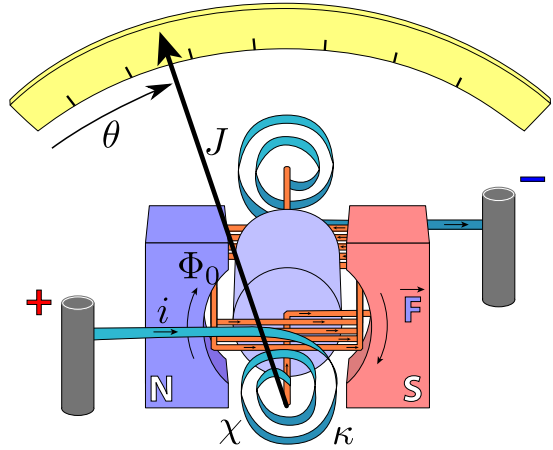}
\end{center}
\caption{{\bf Physical layout of the D'Arsonval galvanometer:~}
We model the essential features of the D'Arsonval meter as: the
rotational moment of inertial of the coil $J$, the torsional spring
constant, $\kappa$, the torsional damping constant, $\chi$ and the
maximum magnetic-flux linked by the coil, $\Phi_0$. We use a linear
model for the stored energy in the coil, $\mathcal{U}_M = \Phi_0 i
{\theta ^ {\left( 0 \right)}}$. The Lagrangian function can be written
in terms of these fundamental parameters. (Adapted from the Wikimedia commons.)}
\label{fig:darsonval_physical}
\end{figure}

The simplest form of electric motor is a piece of wire, moving in a
magnetic field. A short element of wire, $dl$, moving in a magnetic
field will experience Lorenz force of $dF = B \cdot i \cdot dl$,
provided that the wire and the fields are orthogonal.  The modern form
of the moving-coil current meter is the result of a long line of
development, which includes contributions from many people, including
Oersted, Schweigger, Kelvin, D'Arsonval, Weston, and Ayrton. A typical
physical meter is shown in
Figure~\ref{fig:darsonval_physical}.  The meter is carefully designed to
guarantee that the magnetic flux density, ${\bf B}$, is orthogonal to
the moving wires.  We can use the Lorenz force on the wire and the
radius of the motion of the wire, $r$, to calculate the rate of energy
that is transduced per unit angle of motion:
\begin{equation}
  \label{eq:energy_transduction_differential}
  d\mathcal{U}_M = \left( B l r \right) \cdot i  \cdot {d{\theta}}  = \Phi_0 \cdot i  \cdot {d{\theta}} ,
\end{equation}
where $\Phi_0$ is a parameter that represents the construction of the meter.
If we use multiple turns of wire then this simply re-scales the parameter, $\Phi_0$, but does
not alter the basic model. We can integrate Equation~\ref{eq:energy_transduction_differential} to obtain
\begin{equation}
  \label{eq:energy_transduction_integrated}
  \mathcal{U}_M = \Phi_0 \cdot i \cdot \theta ,
\end{equation}
which is the appropriate energy term for the Lagrangian function of a moving, current carrying coil, in a magnetic field.
We note that the energy that can be transduced is unbounded, if the angle, $\theta$, is allowed to increase without bound.
Of course, this is normal for a motor. On practice the angle for the meter cannot increase outside of the range
$-90^{\circ} < \theta < +90^{\circ}$ because the meter does not have a commutator. 
Forces would cease and then change direction at the boundaries.\\

\begin{figure}[!ht]
\begin{center}
\includegraphics[width=0.45\textwidth]{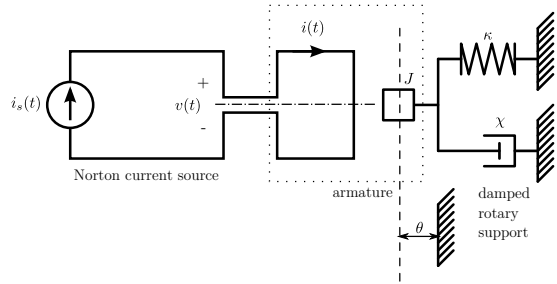}
\end{center}
\caption{{\bf An equivalent electro-mechanical circuit for a D'Arsonval galvanometer:~}
The Lagrangian function for this mechatronic system is written 
in Equation~\ref{eq:darsonval_euler_lagrange}.
The current, $i$, comes from an ideal
current source, so it is essentially a constraint, rather than an independent coordinate. The last term in this
Lagrangian function determines the coupling between the electrical and mechanical aspects of this system.}
\label{fig:darsonval_schematic}
\end{figure}

The moving-coil current meter is shown in Figure~\ref{fig:darsonval_physical}.
The relevant parameters of this physical system
are, the rotational moment of inertial of the coil (including the needle and the physical supports) $J$, the
torsional Hooke's law spring constant, $\kappa$, the torsional damping constant, $\chi$ and the maximum
magnetic-flux linked by the coil, $\Phi_0$, defined earlier. For a coil in free space the stored energy in the coil is given by
$\mathcal{U}_M = \Phi_0 \cdot i \cdot \sin \left( \theta ^ {\left( 0 \right)} \right)$, but for the D'Arsonval and Weston
style of meter the magnetic field is at right angles to a narrow circular air-gap. This makes the magnetic
stored energy function more linear with respect to $\theta$. We can write $\mathcal{U}_M = \Phi_0 \cdot i \cdot {\theta ^
{\left( 0 \right)}}$, as in Equation~\ref{eq:energy_transduction_integrated}. 
These abstract aspects of the meter and their relationships are shown in
Figure~\ref{fig:darsonval_schematic}. There is only one generalised coordinate for this system, $\theta$ and the
Lagrangian function for the linearised system can be written as
\begin{eqnarray}\label{eq:darsonval_lagrangian}
   \mathcal{L} & = & \frac{J}{2} \left( \theta^{\left( 1 \right)} \right)^2         \nonumber \\
               &   & -  \frac{\kappa}{2} \left( \theta^{\left( 0 \right)} \right)^2 \nonumber \\
               &   & + \frac{j}{2} \chi \left( \theta^{\left( 1/2 \right)} \right)^2 \nonumber \\
               &   & + \Phi_0 i {\theta}^{\left( 0 \right)}.
\end{eqnarray}
The last term represents the transduction of energy through the coil.\\

We can apply Equation~\ref{eq:euler_lagrange_general_fractional}
to Equation~\ref{eq:darsonval_lagrangian}
and obtain the equation of motion for the D'Arsonval meter:
\begin{equation}
\label{eq:darsonval_euler_lagrange}
 J \theta^{\left( 2 \right)} + \chi \theta^{\left( 1 \right)} + \kappa  \theta^{\left( 0 \right)} = \Phi_0 i. 
\end{equation}
This example shows that it is possible to model mixed mechanical and electrical (mechatronic) systems using Lagrangian techniques.
Further, we show that the presence of vicious damping is no obstacle to Lagrangian analysis.\\

\section*{Summary and conclusions}
\label{sec:summary}

We have extended the range of applications of Lagrangian analysis, to include non-conservative systems that include dissipative forces.
This has been achieved, even though it contradicts many of the accepted ideas in the current literature.
We have also provided a systematic method of applying an extended type of Lagrangian analysis to non-conservative mechatronic systems.\\

Our approach motivates a number of directions for future work:
\begin{itemize}

\item It is possible to extend Lagrangian techniques to
non-linear dissipative systems, such as memristors or diodes, using
Taylor's theorem, or by using repeated integration by parts, which is
how Cauchy proved Taylor's theorem.

\item If we could extend fractional calculus of variations to include
generalised functions, such as white noise, then we could develop a
fractional Malliavin calculus.  The greater aim is to analyse
electromechanical systems in the presence of noise.  We expect that
this would lead to the solution of the apparent paradoxes of the
Penfield motor~\cite{penfield_1966}, and the Davis electromechanical
capacitor~\cite{davis_2001}.

\item Extremal principles can be used to create a number of
numerical methods, including the method of Rayleigh-Ritz.  It is
possible to simulate mechatronic systems without having to formulate
or solve the ordinary differential equations of motion for those
systems. It is possible to write down the Lagrangian function for a
system, specify any constraints, and then find approximate solutions
using numerical optimisation packages. The authors have had some
success with \texttt{fincon()} in Matlab, and with \texttt{sqp()} in
GNU~Octave. Such methods can be iterative, so an
approximate solution can always be improved, through further iteration

\item It is possible to apply Noether's theorem to determine
the ``constants of the motion'' for quite general systems, including
systems with dissipative elements and dependent sources. These
constants of the motion will be generalised forms of momentum and
energy.  It seems reasonable to suppose that the constants of the
motion could be used to construct Lyapunov functions, which may then
be used to make statements about the stability, or otherwise, of
mechatronic systems.

\item For noisy electrical systems, with many degrees of freedom, it
is of great theoretical interest to write down Liouville's theorem, in
its most general form. The greater aim here is to understand the
thermodynamics of electrical systems.

\item It should be possible to create a time-average Lagrangian
analysis for switched-mode systems. This would be analogous to the
time-averaged state-space models of Middlebrook and
{\'C}uk~\cite{middlebrook_1976}.

\end{itemize}

In summary, we argue that the generalised Lagrangian functions
described in this paper are expected to have impact on  theoretical and
practical applications in electrical and mechanical engineering.\\


\section*{Acknowledgements}
\label{sec:acknowledgements}

The authors thank Chris~Illert for his helpful suggestions. 
He suggested that it was possible to construct a consistent and completely variational approach to dynamical problems. 
He also pointed out that Lagrangian Functions can be imaginary, or complex.
This is discussed in Illert~\cite{illert_1989}, for example.\\

The authors also thank Gareth~Bridges for helping with the proofing of the diagrams, and the choice of symbols and some aspects of notation.\\


\bibliographystyle{plos2009}

\bibliography{F_VACT_201211090842} 

\begin{thebibliography}{10}
\providecommand{\url}[1]{\texttt{#1}}
\providecommand{\urlprefix}{URL }
\expandafter\ifx\csname urlstyle\endcsname\relax
  \providecommand{\doi}[1]{doi:\discretionary{}{}{}#1}\else
  \providecommand{\doi}{doi:\discretionary{}{}{}\begingroup
  \urlstyle{rm}\Url}\fi
\providecommand{\bibAnnoteFile}[1]{%
  \IfFileExists{#1}{\begin{quotation}\noindent\textsc{Key:} #1\\
  \textsc{Annotation:}\ \input{#1}\end{quotation}}{}}
\providecommand{\bibAnnote}[2]{%
  \begin{quotation}\noindent\textsc{Key:} #1\\
  \textsc{Annotation:}\ #2\end{quotation}}
\providecommand{\eprint}[2][]{\url{#2}}

\bibitem{feynman_1963}
Feynman RP, Leighton RB, Sands M (1963) {T}he {F}eynman {L}ectures on
  {P}hysics, volume~1.
\newblock Reading, Massachusetts: Addison-Wesley.
\bibAnnoteFile{feynman_1963}

\bibitem{lanczos_1949}
Lanczos C (1949) The {V}ariational {P}rinciples of {M}echanics.
\newblock {D}over~Publications,~Inc.
\bibAnnoteFile{lanczos_1949}

\bibitem{karnopp_2000}
Karnopp DC, Margolis DL, Rosenberg RC (2000) {S}ystem {D}ynamics, {M}odelling
  and {S}imulation of {M}echatronic {S}ystems.
\newblock New York: John Wiley and Sons Inc.
\bibAnnoteFile{karnopp_2000}

\bibitem{penrose_2005}
Penrose R (2005) {T}he {R}oad to {R}eality.
\newblock London: {V}intage {B}ooks.
\bibAnnoteFile{penrose_2005}

\bibitem{post_1962}
Post EJ (1962) {F}ormal {S}tructure of {E}lectromagnetics.
\newblock New York: {D}over~{P}ublications~{I}nc., {D}over~1997 edition.
\bibAnnoteFile{post_1962}

\bibitem{penfield_1966}
Penfield~Jr P (1966) Unresolved paradox in circuit theory.
\newblock Proceedings of the IEEE 54: 1200--1201.
\bibAnnoteFile{penfield_1966}

\bibitem{melia_2001}
Melia F (2001) {E}lectrodynamics.
\newblock Chicago: {U}niversity of {C}hicago {P}ress.
\bibAnnoteFile{melia_2001}

\bibitem{riewe_1996}
Riewe F (1995) {N}onconservative {L}agrangian and {H}amiltonian mechanics.
\newblock Physical Review E 52: 1890--1899.
\bibAnnoteFile{riewe_1996}

\bibitem{riewe_1997}
{R}iewe F (1997) Mechanics with fractional derivatives.
\newblock Physical Review E 55: 3581--3592.
\bibAnnoteFile{riewe_1997}

\bibitem{gelfand_1963}
{Gel{'}fand} IM, Fomin SV (1963) {C}alculus of {V}ariations.
\newblock New York: {D}over~{P}ublications~{I}nc., {D}over~1991 edition.
\bibAnnoteFile{gelfand_1963}

\bibitem{smith_1974}
Smith DR (1974) Variational Methods in Optimization.
\newblock New York: {D}over~{P}ublications~{I}nc., {D}over~1998 edition.
\bibAnnoteFile{smith_1974}

\bibitem{agrawal_2001}
Agrawal OP (2001) {F}ormulation of {E}uler-{L}agrange equations for fractional
  variational problems.
\newblock J~Math~Anal~Appl 272: 368--379.
\bibAnnoteFile{agrawal_2001}

\bibitem{rabei_2004}
Rabei EM, Alhalholy TS (2004) On {H}amiltonian formulation of non-conservative
  systems.
\newblock Turk J Phys 28: 213--221.
\bibAnnoteFile{rabei_2004}

\bibitem{frederico_2007}
Frederico GSF, Torres DFM (2007) A formulation of {N}oether's theorem for
  fractional problems of the calculus of variations.
\newblock Math~Anal~Appl 334: 834--846.
\bibAnnoteFile{frederico_2007}

\bibitem{musielak_2007}
Musielak ZE (2008) {S}tandard and non-standard {L}agrangians for dissipative
  systems with variable coefficients.
\newblock Journal of Physics A: Mathematical and Theoretical 41: 055205.
\bibAnnoteFile{musielak_2007}

\bibitem{elnabulsi_2008}
El-Nabulsi RA (2008) Fractional action-like variational problems.
\newblock Journal of Mathematical Physics 49: 1--7.
\bibAnnoteFile{elnabulsi_2008}

\bibitem{almeida_2009}
Almeida R, Torres DFM (2009) Calculus of variations with fractional derivatives
  and fractional integrals.
\newblock Applied Mathematics Letters 22: 1816--1820.
\bibAnnoteFile{almeida_2009}

\bibitem{almeida_2010}
Almeida R, Malinowska AB, Torres DFM (2010) A fractional calculus of variations
  for multiple integrals with application to a vibrating string.
\newblock Journal of Mathematical Physics 51: pp1--12.
\bibAnnoteFile{almeida_2010}

\bibitem{oldham_1974}
Oldham KB, Spanier J (1974) {T}he {F}ractional {C}alculus.
\newblock New York: {D}over~{P}ublications~{I}nc., {D}over~1974 edition.
\bibAnnoteFile{oldham_1974}

\bibitem{weisstein_1999}
Weisstein EW (1999) CRC Concise Encyclopedia of Mathematics.
\newblock New York: CRC Press.
\bibAnnoteFile{weisstein_1999}

\bibitem{lamb_1946}
Lamb H (1946) {D}ynamics.
\newblock Cambridge: Cambridge University Press.
\bibAnnoteFile{lamb_1946}

\bibitem{goldstein_1950}
Goldstein H (1950) {C}lassical~{M}echanics.
\newblock Reading,~Massachusetts: Addison-Wesley.
\bibAnnoteFile{goldstein_1950}

\bibitem{mccuskey_1959}
McCuskey SW (1959) {A}n {I}ntroduction to {A}dvanced {D}ynamics.
\newblock Reading,~Massachusetts: Addison-Wesley.
\bibAnnoteFile{mccuskey_1959}

\bibitem{resnick_1960}
Resnick R, Halliday D (1960) {P}hysics.
\newblock New~York: John~Wiley~{\&}~Sons~Inc.
\bibAnnoteFile{resnick_1960}

\bibitem{wylie_1960}
{Wylie,~Jr} CR (1960) {A}dvanced {E}ngineering {M}athematics.
\newblock New~York: McGraw-Hill Book Co.
\bibAnnoteFile{wylie_1960}

\bibitem{fowles_1962}
Fowles GR (1962) {A}nalytical {M}echanics.
\newblock New York: Holt, Rrinhart and Winston.
\bibAnnoteFile{fowles_1962}

\bibitem{rabenstein_1972}
Rabenstein AL (1972) {I}ntroduction to {O}rdinary {D}ifferential {E}quations.
\newblock New~York: Academic~Press.
\bibAnnoteFile{rabenstein_1972}

\bibitem{lomen_1988}
Lomen D, Mark J (1988) {D}ifferential {E}quations.
\newblock Englewood~Cliffs,~NJ: Prentice~Hall.
\bibAnnoteFile{lomen_1988}

\bibitem{giesecke_1980}
Giesecke F, Mitchell A, Spencer HC, Hill IL, JTDygdon (1980) {T}echnical
  {D}rawing.
\newblock New~York: Macmillan~Publishing~Co. Inc.
\bibAnnoteFile{giesecke_1980}

\bibitem{williams_1996}
{{W}illiams,~Jr} JH (1996) {F}undamentals of {A}pplied {D}ynamics.
\newblock New York: {J}ohn~{W}iley~{\&}~{S}ons~{I}nc.
\bibAnnoteFile{williams_1996}

\bibitem{reif_1965}
Reif F (1965) {F}undamentals of {S}tatistical and {T}hermal {P}hysics.
\newblock Singapore: McGraw~Hill, {I}nternational~{E}d.~1985 edition.
\bibAnnoteFile{reif_1965}

\bibitem{wannier_1966}
Wannier GH (1966) {S}tatistical {P}hysics.
\newblock New York: {D}over Publications Inc., {D}over 1987 edition.
\bibAnnoteFile{wannier_1966}

\bibitem{allison_2009}
Allison A (2009) {A}spects of {S}tochastic {C}ontrol and {S}witching: from
  {{P}arrondo{'}s} games to {E}lectrical {C}ircuits.
\newblock Ph.D. thesis, Electrical and Electronic Engineering, University of
  Adelaide, Adelaide.
\bibAnnoteFile{allison_2009}

\bibitem{illert_1989}
Illert C (1989) Formulation and solution of the classical seashell problem
  ii.-tubular three-dimensional seashell surfaces.
\newblock Il Nuovo Cimento D 11: 761--780.
\bibAnnoteFile{illert_1989}

\bibitem{jaynes_1980}
Jaynes ET (1980) The minimum entropy production principle.
\newblock Ann~Rev~Phys~Chem 31: 579--601.
\bibAnnoteFile{jaynes_1980}

\bibitem{kirchhoff_1848}
Kirchhoff G (1848) {U}eber die {A}nwendbarkeit der {F}ormeln f{\"{u}}r die
  {I}ntensitäten der galvanischen {S}tr{\"{o}}me in einem {S}ysteme linearer
  {L}eiter auf {S}ysteme, die zum {T}heil aus nicht linearen leitern bestehen.
\newblock Ann Phys 151: 189–-205.
\bibAnnoteFile{kirchhoff_1848}

\bibitem{davis_2001}
Davis BR, Abbott D, Parrondo JMR (2001) Thermodynamic energy exchange in a
  moving plate capacitor.
\newblock Chaos 11: 747--754.
\bibAnnoteFile{davis_2001}

\bibitem{middlebrook_1976}
Middlebrook RD, {\'C}uk SA (1976) A general unified approach to modelling
  switching converter power stages.
\newblock In: IEEE Power Electronics Specialist's Conf Rec. pp. 18--34.
\bibAnnoteFile{middlebrook_1976}

\end{thebibliography}


\end{document}